\documentclass[epjST]{svjour}

\usepackage{graphicx}
\usepackage{amsmath}
\usepackage{amssymb}

\graphicspath{{./figures/}}

\newcommand{\abs}[1]{\left\vert #1 \right\vert}
\renewcommand{\vec}[1]{\boldsymbol{#1}}
\renewcommand{\Re}[1]{{\rm Re}\left[#1\right]}
\renewcommand{\Im}[1]{{\rm Im}\left[#1\right]}

\def \a  {\alpha}
\def \D  {\Delta}
\def \g  {\gamma}
\def \vp {\varphi}
\def \Vp {\Phi}
\def \w  {\omega}
\def \e  {\varepsilon}
\def \t  {\theta}
\def \p  {\psi}
\def \P  {\Psi}
\def \b  {\beta}
\def \t  {\theta}
\def \T  {\Theta}
\def \r  {\rho}
\def \s  {\sigma}
\def \n  {\nu}

\title{Using Phase Dynamics to Study Partial Synchrony: Three Examples}
\author{Erik Teichmann\thanks{\email{kontakt.teichmann@gmail.com}}}
\institute{Institute of Physics and Astronomy, University of Potsdam,
Karl-Liebknecht-Str. 24/25, 14476 Potsdam-Golm, Germany}

\abstract{Partial synchronous states appear between full synchrony and
asynchrony and exhibit many interesting properties.  Most frequently, these
states are studied within the framework of phase approximation.  The latter is
used ubiquitously to analyze coupled oscillatory systems.  Typically, the phase
dynamics description is obtained in the weak coupling limit, i.e., in the
first-order in the coupling strength.  The extension beyond the first-order
represents an unsolved problem and is an active area of research.  In this
paper, three partially synchronous states are investigated and presented in
order of increasing complexity.  First, the usage of the phase response curve
for the description of macroscopic oscillators is analyzed.  To achieve this,
the response of the mean-field oscillations in a model of all-to-all coupled
limit-cycle oscillators to pulse stimulation is measured.  The next part treats
a two-group Kuramoto model, where the interaction of one attractive and one
repulsive group results in an interesting solitary state, situated between full
synchrony and self-consistent partial synchrony.  In the last part, the phase
dynamics of a relatively simple system of three Stuart-Landau oscillators are
extended beyond the weak coupling limit.  The resulting model contains triplet
terms in the high-order phase approximation, though the structural connections
are only pairwise.  Finally, the scaling of the new terms with the coupling is
analyzed.}

\begin{document}
\maketitle

\section{Introduction}
Some of the first observations of the phenomenon of synchronization have been
made in the late 17th century. The Dutch physician Kaempfer observed a swarm of
fireflies in Asia and noted their rhythmic flashing; their lights appeared in a
regular interval all over the whole swarm~\cite{kaempfer1906history}. Some years
earlier, the Dutch physicist Huygens already noted that two pendulum clocks,
fastened to the same beam, always swung in opposite directions, regardless of
where and how he released them~\cite{pikovsky2003synchronization}.

Despite essential progress, synchronization remains a topic of active
research. In particular, many studies are devoted to the investigation of
different partially synchronous states.  Partial synchrony describes the state
between full synchrony and asynchrony, where not all phases or frequencies are
equal. In most systems, partial synchrony exists in the biggest part of the
parameter space, making it an important topic to study.  Some interesting
realizations of partial synchronous states are the
chimera~\cite{kuramoto2002coexistence},
Bellerophon~\cite{qiu2016synchronization}, traveling
wave~\cite{hooper1988travelling}, or solitary
state~\cite{maistrenko2014solitary}.

The study of self-sustained oscillations has become an important tool to describe
phenomenons as diverse as the common movement of pedestrians on a
bridge~\cite{strogatz2005theoretical}, the heart beat~\cite{pol1928heartbeat} or
the motion of a fish swarm~\cite{ashraf2016synchronization} and their
synchronization properties. The application to neuroscience, where oscillatory
behavior determines the dynamics in the
brain~\cite{breakspear2010generative,cumin2007generalising,tateno2007phase}, is
of special interest.

Self-sustained oscillators are well understood, but a quantitative analysis of
systems of coupled oscillators is generally hard.  The description of
oscillatory systems can consist of numerous coupled differential equations
containing nonlinear terms and only allows for approximate or qualitative
analysis in most cases. One way to reduce the complexity is phase reduction,
which describes every single oscillatory unit with one one-dimensional variable,
thereby reducing the problem's dimensionality.

The phase reduction of a single oscillator is known analytically only for a
few types.  Even for these types of oscillators, coupled units' dynamics are
typically only available in the weak coupling limit, i.e., in the first order of
the coupling strength. Methods for finding the phase dynamics for stronger
couplings are either restricted to coupling functions with a specific
property~\cite{kurebayashi2013phase} or to specific
systems~\cite{leon2019phase}.

While the first-order phase approximation of pairwise coupled oscillators yields
only terms depending on two phases, beyond the weak coupling limit, generally
terms depending on several phases appear. Some of these new terms are triplet
terms or non-structural terms, i.e., connections not present in the coupling
scheme~\cite{matheny2019exotic,leon2019phase}. These are often reconstructed
numerically but are seen as spurious terms or correlations.

In this minireview, I summarize the current state of research for my Ph.D.
thesis by analyzing three cases of partial synchrony with the help of phase
dynamics in increasing levels of complexity.  In the first level in
section~\ref{sec:prc}, the macroscopic phase dynamics of a complex mean-field
oscillator, generated by the collective dynamics of the single units, is
investigated. The novel results about the direct extension of the phase from a
single oscillator to an ensemble of oscillators are shown via the collective
phase response curve (PRC) for the mean-field. It measures the mean-field's
reaction to a perturbation of the oscillators in the first order of the
perturbation strength and is applied to coupled Rayleigh oscillators. In the
second level, a more sophisticated approach, the weak coupling limit, is used to
describe the phases of the units in the first order of the coupling strength. In
the Kuramoto model~\cite{kuramoto1984chemical}, this approach is used widely to
investigate synchronization. Here it is applied to two groups of oscillators,
one attractive and one repulsive, to investigate an interesting solitary state
in section~\ref{sec:kuramoto}, and summarizes the
paper~\cite{teichmann2019solitary}. Finally, in section~\ref{sec:sl}, the phase
reduction is expanded past the weak coupling limit, as the third level of
complexity, in a system of Stuart-Landau oscillators using a perturbation
method, as described in~\cite{gengel2020high}. This approach reveals additional
terms in the phase model. For example, triplet terms appear for pairwise coupled
oscillators.

\section{Phase Dynamics}
Phase dynamics are defined for an oscillatory dynamical system of arbitrary
dimension
\begin{equation}
  \frac{d \vec{y}}{dt} = \vec{f}(\vec{y})
  \label{eq:dynamical_system}
\end{equation}
with a stable limit cycle of period $T$ in the state space ${\vec{Y}(t + T) =
\vec{Y}(t)}$. On this limit cycle, and in its basin of attraction, a phase ${\vp
= \Vp(\vec{y})}$ can be defined that identifies the state uniquely and grows
uniformly in time by
\begin{equation}
  \dot{\vp} = \w = \frac{\partial \Vp}{\partial \vec{y}} \vec{f}(\vec{y})
  \label{eq:phase_definition}
\end{equation}
with a frequency of ${\w = \frac{2 \pi}{T}}$~\cite{pikovsky2003synchronization}.
This phase reduces the dimensionality of the dynamics to just one. For most
types of oscillators the phase dynamics are not known analytically and have to
be reconstructed numerically.

In Eq.~\eqref{eq:phase_definition} only the autonomous dynamics are considered.
In the case of a weak perturbation the phase instead evolves according to the
Winfree equation~\cite{winfree1980geometry} in the first order of the
perturbation strength $P$.  With the PRC $\D$ and the perturbation $p$ it reads
\begin{equation}
  \dot{\vp} = \w + \D(\vp) p(t) + \mathcal{O}(P^2)\; .
  \label{eq:prc}
\end{equation}
The PRC $\D$ describes the phase change depending on the current phase. It is
also an important tool in describing the system beyond the phase dynamics, as
its form gives information about the stability and synchronization properties of
the system~\cite{achuthan2009phasea}.

\subsection{Phase Dynamics of a Macroscopic Oscillator - Rayleigh Model}
\label{sec:prc}

A natural extension of the PRC in the case of multiple oscillators is the
collective PRC, which describes the reaction of a whole ensemble of $N$
oscillators to a perturbation. Their dynamics are measured by their average, the
mean-field ${\vec{Z} = 1/N \sum_j \vec{y}_j}$.  For a strong enough coupling,
the mean-field will also move on a limit cycle, and a perturbation of the
oscillators will lead to a deviation from its stable trajectory. This means the
mean-field can be seen as a complex oscillator, as demonstrated for the brain
rhythm in experiments with rats~\cite{velazquez2015epileptic}. From the reaction
of the mean-field oscillator, it is even possible to gain information about the
PRC of single oscillators~\cite{wilson2015determining}.

Whereas the PRC $\D$ describes the complete reaction to a perturbation, the
collective PRC can be split into two parts. The prompt PRC $\D_0$ is the
immediate reaction of the mean-field and the relaxation PRC $\D_r$ the part that
describes the relaxation of the oscillators after the perturbation and the
resulting change in their distribution. Together these form the final PRC
$\D_f$~\cite{levnajic2010phase,hannay2015collective}. Formally this can be
written as
\begin{align}
  \D_0(\vp_0) & = \bar{\vp}_0 - \vp_0 \; , \\
  \D_f(\vp_0) & = \lim_{t \to \infty} (\bar{\vp}(t - \tau) - \vp(t - \tau))
    = \D_0(\vp_0) + \lim_{t \to \infty} \D_r(\vp_0, t - \tau) \; ,
  \label{eq:final_prc}
\end{align}
where $\tau$ is the time of the perturbation, $\vp_0$ is the phase at the time
of perturbation of the unperturbed system and $\bar{\vp}$ the phase in the
perturbed system. Because there does not exist a general way to describe the
dynamics of the mean-field, the collective PRC and its parts have to be
calculated numerically.

Numerically the collective PRC is found by measuring the change in the $k$-th
period of the mean-field after the perturbation $\bar{T}_k$ and the unperturbed
period $T$ as
\begin{equation}
  \D_N(\vp_0) = 2 \pi \sum_{j=1}^N \frac{T - \bar{T}_k}{T} \; .
\end{equation}
After a sufficiently long time this will approach the final PRC in
Eq.~\eqref{eq:final_prc}.

Consider a system of $N$ coupled Rayleigh oscillators~\cite{rayleigh1945theory}
with the dynamics
\begin{equation}
  \ddot{x}_k - \eta (1 - \dot{x}_k^2) \dot{x}_k + \w_k^2 x_k = \e (\dot{X} - \dot{x}_k) \; .
  \label{eq:rayleigh_ode}
\end{equation}
The $\eta$ is a nonlinearity parameter, $\e$ the coupling strength and ${\dot{X}
= 1/N \sum_j \dot{x}_j}$. The natural frequencies $\w_k$ are distributed
according to a Gaussian distribution with mean $1$ and standard deviation
$0.01$. For $\eta = 6$ and ${N = 500}$ this system has a stable limit cycle of
the mean-field for ${\e \gtrapprox 0.134}$ and is in the partial synchronous
regime.

\begin{figure}
  \includegraphics{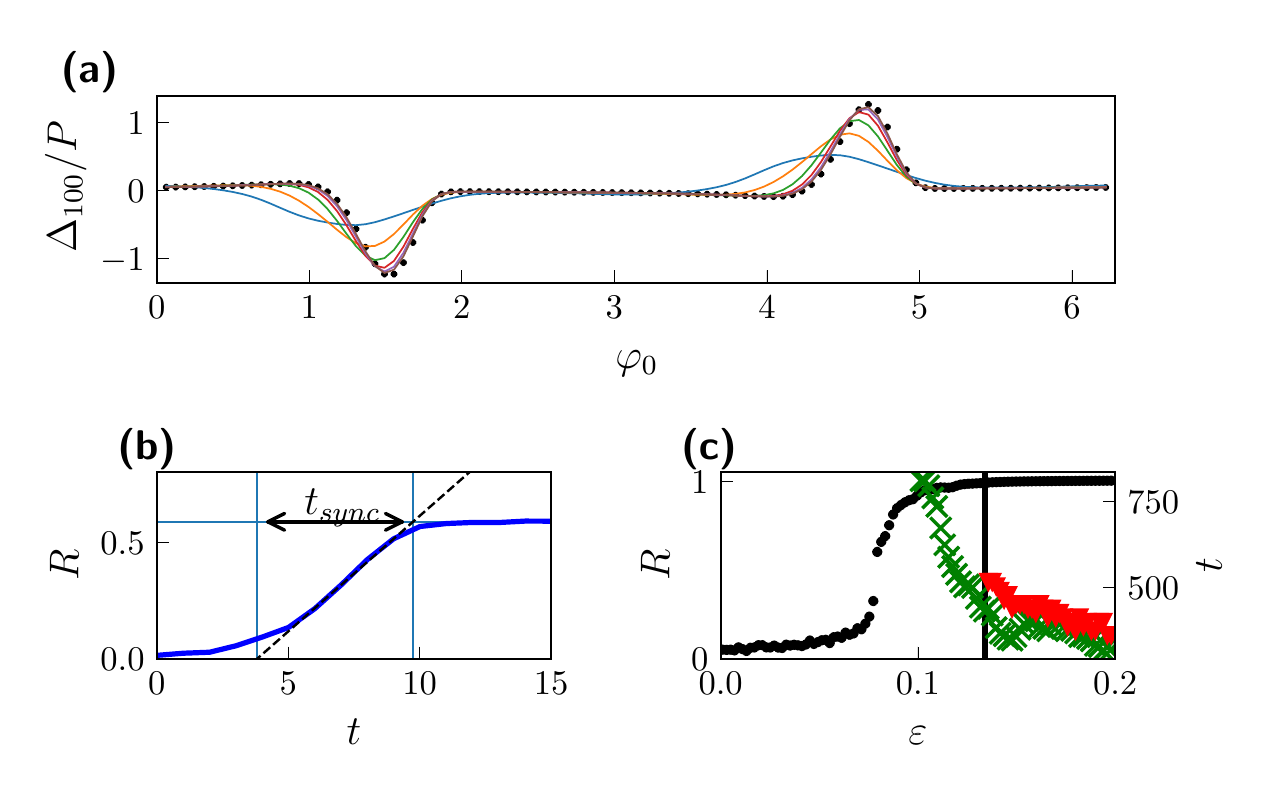}
  \caption{The collective PRC $\D_{100}$ for the Rayleigh model in
  Eq.~\eqref{eq:rayleigh_ode} with perturbation strength $P$ is shown in (a).
  The black dots mark the PRC of the single oscillator and the lines the
  numerically measured values for $\e \in \{0.134, 0.336, 0.538, 0.74, 0.942\}$,
  where the collective PRC approaches the single oscillator PRC with increasing
  $\e$. In (b) the method for the calculation of the synchronization time
  $t_{sync}$ is visualized. The timescales for the synchronization and the
  relaxation are shown in (c). The black dots denote the magnitude of the limit
  cycle at the crossing of the positive $\dot{x}$-axis, the green crosses
  $t_{sync}$ and the red triangles $t_{relax}$. The necessary $\e$ for a stable
  limit cycle is marked with a black line.}
  \label{fig:rayleigh_prc}
\end{figure}

The oscillators are perturbed simultaneously in the $\dot{x}$-direction with a
small enough strength $P$, such that the collective PRC scales linearly with $P$
and Eq.~\eqref{eq:prc} is valid.  The resulting response of the mean-field is
shown in Fig.~\ref{fig:rayleigh_prc}(a). While the collective PRC is flat for
small $\e$, it approaches the PRC of a single oscillator with increasing
coupling strength.  The approach to the PRC of a single oscillator is the
expected outcome, as in the case of $\e \to \infty$, the oscillators will be
fully synchronized and behave like a single unit.

An important property of the collective PRC is the needed relaxation time
$t_{relax}$, as the application to real-world noisy systems becomes impossible
if it is too long. When the oscillators are perturbed strongly before they fully
relax, then the mean-field will not reach its limit cycle, and the phase
dynamics in Eq.~\eqref{eq:prc} are not applicable. To have a comparable
timescale consider the synchronization time $t_{sync}$ in
Fig.~\ref{fig:rayleigh_prc}(b).  It is measured as a linear approximation of the
time needed to reach the stable distribution of oscillators, given by $R =
\abs{\vec{Z}} \approx const$, from a splay state, where all oscillators are
distributed uniformly in the phase. The slope for the linear approximation is
chosen as the value at the inflection point, i.e., the biggest $\dot{R}$. A
comparison of the time scales in Fig.~\ref{fig:rayleigh_prc}(c) shows that
$t_{relax}$ (at about $40T$) is longer than $t_{sync}$. This suggests a very
weak attraction to the stable distribution. The collective PRC can thus be only
applied in an approximate sense.

\section{Phase Dynamics of Coupled Oscillators}

After investigating the phase dynamics under the influence of a perturbation,
the next step of abstraction is the application of phase dynamics to coupled
oscillators.  Instead of the mean-field, the interest lies now on the single
oscillatory units and their phases. The dynamics of coupled oscillators with
coupling function $\vec{G}_k$ for oscillator $k$ and coupling strength $\e$ are
\begin{equation}
  \frac{d \vec{y}_k}{dt} = \vec{f}_k(\vec{y}_k)
    + \e \vec{G}_k(\vec{y}_1, \vec{y}_2, \ldots) \; .
\end{equation}
The phase dynamics for such a system are given analogous to
Eq.~\eqref{eq:phase_definition} with  ${\vp_k = \Vp_k(\vec{y}_k)}$ as
\begin{equation}
  \dot{\vp}_k = \w_k
  + \e \frac{\partial \Vp_k}{\partial \vec{y}_k}
  \vec{G}_k (\vec{y}_1, \vec{y}_2, \ldots) \; .
\end{equation}
In this case, the phase does not grow uniformly. While it could be found on the
limit cycle by virtue of the period, now the phases of all points in the state
space have to be known to solve the coupling term.  With the approximation of
weak coupling, the dynamics stay close to the limit cycle of the uncoupled unit
and the phases in this region are known, so the coupling function can be written
in terms of these phases
\begin{equation}
  \dot{\vp}_k = \w_k + \e G_k(\vp_k, \vp_1, \vp_2, \ldots)
    + \mathcal{O}(\e^2) \; .
  \label{eq:phase_first_order}
\end{equation}
Higher-order approximations have to be considered to extend the phase dynamics
farther past the limit cycle, for example, when considering higher coupling
strengths.

Some methods for analytical phase reduction of higher-order consider systems
with a separation of timescales~\cite{kurebayashi2013phase,pyragas2015phase} or
use isostable coordinates~\cite{wilson2016isostable}. A general method that only
works for systems with a known phase in the autonomous systems uses a
perturbation Ansatz on the isochrones~\cite{leon2019phase}. To reconstruct the
phase dynamics numerically, one typically uses a Fourier series to represent the
coupling function and then fits the coefficients using, e.g., a multiple
shot~\cite{tokuda2007inferring} or Bayesian
methods~\cite{stankovski2016alterations}. Other numerical reductions include
reconstruction of the phase from a polar phase that is determined by a Hilbert
transformation~\cite{kralemann2008phase,kralemann2011reconstructing} or directly
with an iterative Hilbert transform~\cite{gengel2019phase}. Some improvements
can be made by using absolute phases in the coupling function instead of phase
differences~\cite{blaha2011reconstruction} or considering triplets of
oscillators to find structural connectivity~\cite{kralemann2014reconstructing}.

\subsection{Weakly Coupled Oscillators - Kuramoto Model}
\label{sec:kuramoto}
The Kuramoto model~\cite{acebron2005kuramoto,pikovsky2015dynamics} describes a
weakly pairwise all-to-all coupled system of oscillators. The first order
approximation in Eq.~\eqref{eq:phase_first_order} is valid and the dynamics can
be written as
\begin{equation}
  \dot{\vp}_k = \w_k + \frac{\e}{N} \sum_{j = 1}^N \sin(\vp_j - \vp_k + \a_k) \; .
  \label{eq:kuramoto}
\end{equation}
The $\w_k$ are the natural frequencies and the $\a_k$ the phase shift
parameters.

By using the complex mean-field $Z = R e^{i \t} = 1/N \sum_j e^{i \vp_j}$ the
system can be reduced to
\begin{equation}
  \dot{\vp}_k = \w_k + \e R \sin(\t - \vp_k + \a_k) \; .
\end{equation}
The order parameter $R$ describes the degree of synchronization, $R = 1$ denotes
full synchrony and $R = 0$ is an indicator for incoherence or antisymmetry.

One of the most important properties of this model is the analytical
solvability. In the case of identical oscillators, the powerful
Watanabe-Strogatz (WS) theory can be
used~\cite{watanabe1993integrability,watanabe1994constants}.  It allows for the
reduction of the dynamics from $N$ dimensions to just $3$ and $N-3$ constants of
motion. The remaining degrees of freedom are three variables $\r$, $\P$ and
$\T$. Between them and the mean-field exists a correspondence, although they are
not equal.  Generally, $\r$ and $\T$ are close to $R$ and $\t$, while the final
variable $\P$ can be seen as a measure of the system's clustering. The
correspondence between the degrees of freedom and the mean-field becomes even
bigger in the thermodynamic limit when the dynamics move on the Ott-Antonsen
manifold~\cite{ott2008low,ott2009long} and reduced to just the mean-field
variables, $R$ and $\t$. The Ott-Antonsen manifold can only be seen as an
asymptotic solution, as the transients can be very
long~\cite{pikovsky2011dynamics,mirollo2012asymptotic}.

When investigating multiple ($M$) groups of oscillators, the Kuramoto model can
be extended to the M-Kuramoto model. The groups interact with different coupling
strength and may have phase shifts,
\begin{equation}
  \dot{\vp}_k^{\s} = \w_k^{\s}
    + \sum_{\s' = 1}^M \frac{\e_{\s, \s'}}{N}
    \sum_{j = 1}^{N_{\s'}} \sin(\vp_j^{\s'} - \vp_k^{\s} + \a_{\s, \s'}) \; .
\end{equation}
Here $\s$ and $\s'$ denote the different groups and $\e_{\s, \s'}$ and $\a_{\s,
\s'}$ are the coupling strengths and phase shifts between groups $\s$ and $\s'$,
respectively.

\begin{figure}
  \centering
  \includegraphics{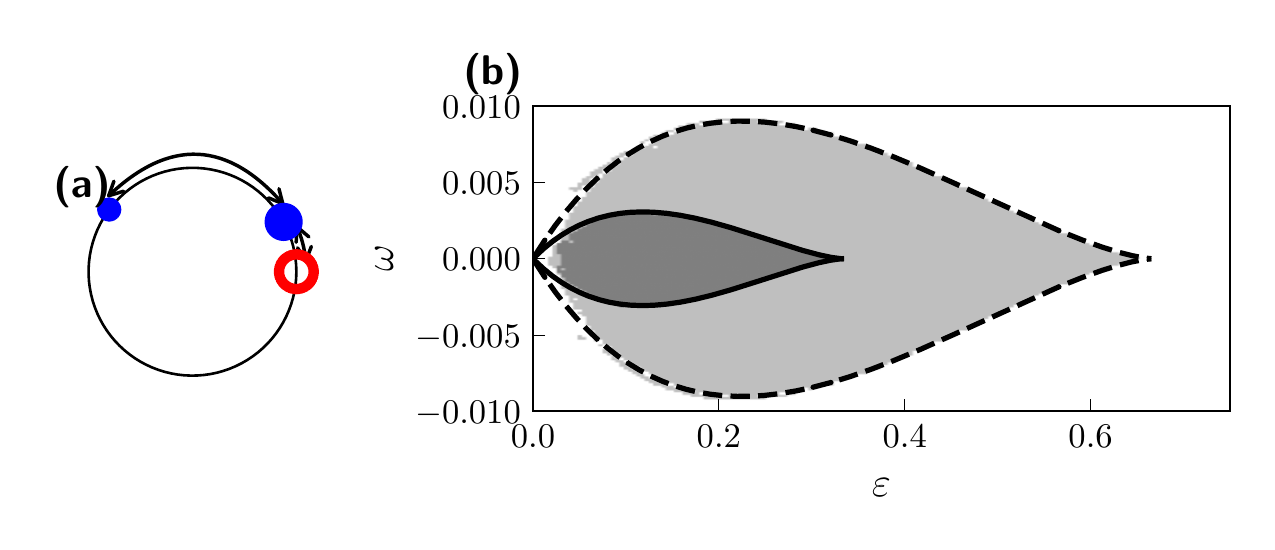}
  \caption{The solitary state in an M-Kuramoto model is visualized in (a). The
  red ring shows the attractive cluster, the big blue circle the repulsive
  cluster and the small blue circle the solitary oscillator. In (b) the
  existence of the solitary state is plotted. The dashed black line shows the
  analytical prediction for $N_a = N_r = 5$ and the light gray area the
  numerical observation. The solid black line and dark gray area show the same,
  but for $N_a = N_r =8$.}
  \label{fig:solitary_state}
\end{figure}

Consider the system with ${\e_{\s, \s'} = \e_{\s'}}$ and ${\a_{\s, \s'} =
\a_{\s'}}$ and two groups of identical oscillators, one attractive $\vp^a$ and
one repulsive $\vp^r$
\begin{equation}
  \begin{aligned}
    \dot{\vp}_k^a & = \w_a
      + \frac{1}{N} \sum_{j = 1}^{N_a} \sin(\vp_j^a - \vp_k^a + \a_a)
      - \frac{1 + \e}{N} \sum_{j = 1}^{N_r} \sin(\vp_j^r - \vp_k^a + \a_r) \\
    \dot{\vp}_k^r & = \w_r
      + \frac{1}{N} \sum_{j = 1}^{N_a} \sin(\vp_j^a - \vp_k^r + \a_a)
      - \frac{1 + \e}{N} \sum_{j = 1}^{N_r} \sin(\vp_j^r - \vp_k^r + \a_r) \; ,
  \end{aligned}
  \label{eq:m_kuramoto_ode}
\end{equation}
where the time was rescaled such that the attractive group has coupling strength
$1$ and the repulsive ${-(1 + \e)}$. Then $\e$ no longer measures the coupling
strength, but the excess of repulsive coupling. These equations can be
reduced further by introducing the mean-fields of the groups ${Z_{a, r} =
1/N_{a,r} \sum e^{i \vp_{a,r}}}$, as before, and a common forcing
\begin{equation}
  H = \frac{N_a}{N} e^{i \a_a} Z_a - \frac{N_r}{N} (1 + \e) e^{i \a_r} Z_r \; .
\end{equation}
The reduced equations are
\begin{equation}
  \begin{aligned}
    \dot{\vp}_a & = \w_a + \Im{H e^{-i \vp_a}} \\
    \dot{\vp}_r & = \w_r + \Im{H e^{-i \vp_r}}
  \end{aligned}
\end{equation}
and allow for the application of the WS theory to the
system~\cite{pikovsky2008partially}.

In the system without natural frequencies ${\w_k^{\s} = 0}$ there exists a
peculiar solitary state~\cite{maistrenko2014solitary}, see
Fig.~\ref{fig:solitary_state}(a).  When both groups have the same size ${N_a =
N_r}$, the attractive group, and all the repulsive oscillators, except for one,
will cluster at the same point. The remaining repulsive oscillator will be
phase-shifted by $\pi$.  In the case of ${\a_{\s'} = 0}$, the state exists for
all values of $N$ in some region of $\e$ (which shrinks for increasing $N$), but
it does not have full measure, i.e., some initial conditions may lead to a
different state.  For ${\a_{\s'} \neq 0}$, the state only exists up to a
critical system size and without full measure. The state's existence is also
predicted by the WS theory and is the only allowed clustered state, except for
full synchrony.

Since the solitary state is a rather new discovery, there are slightly different
definitions of it. The most popular one is that in a system with a
natural order of the oscillators, e.g., ordered by their natural frequency, some
single oscillators behave differently than the rest of the population and, more
importantly, than their immediate neighbors. This is the main difference to a
chimera state, where a whole subpopulation exhibits a different behavior.
Kapitaniak et al. observed such a state in Ref.~\cite{kapitaniak2014imperfect}
with metronomes coupled in a ring to their nearest neighbor and second nearest
neighbor. In such a configuration, some single oscillators' phases will
differ from the rest of their synchronized neighbors. Another observation was
made in Ref.~\cite{hizanidis2016chimera}, where superconducting quantum
interference devices were placed in a one-dimensional array and coupled
magnetically. In this case, the solitary oscillators exhibit a higher amplitude
than the rest of the population.

The solitary state has also been found numerically in simulations of coupled
Lorenz oscillators~\cite{shepelev2018chimera} and in a ring of coupled
Stuart-Landau oscillators with symmetry breaking attractive and repulsive
long-range coupling~\cite{sathiyadevi2019long}. The observation of the existence
in multiplex networks of FitzHugh-Nagumo oscillators coupled in rings with a
small mismatch in the intra-layer couplings~\cite{mikhaylenko2019weak}, allows
even for controlling strategies to tune the dynamics in, e.g., neural networks.

While the existence has been shown in many different systems, there are only a
few results in investigating its emergence and stability. In a Kuramoto model
with inertia, the solitary state arises from a homoclinic bifurcation and
persists even in the thermodynamic limit~\cite{jaros2018solitary}, in contrast
to it being a finite size effect in the M-Kuramoto model with attractive and
repulsive interaction.  Aside from differential equations, the solitary state
has also been found in coupled maps.  Multiplex network of non-locally coupled
maps with a singular hyperbolic attractor exhibit solitary states in their
transition from coherence to
incoherence~\cite{rybalova2017transition,semenova2017coherenceincoherence}.
During the transition, more and more solitary oscillators appear, growing almost
linearly with the decrease in the coupling strength. This is the result of an
increase of the size of the basin of attraction of the solitary set with a
decrease in the coupling, as more random initial conditions lie in this
basin~\cite{semenova2018mechanism}. To also induce solitary states in maps with
nonhyperbolic attractors, a multiplicative noise can be added to the coupling
constant, thus also showing the existence of the solitary state in a noisy
system~\cite{rybalova2018mechanism}.

These solitary state can consist of multiple solitary oscillators, but from
here on, a solitary state is defined more narrowly, such that a single oscillator
shows a different behavior than the rest of the population.

In Ref.~\cite{teichmann2019solitary} we extend the M-Kuramoto model from
Ref.~\cite{maistrenko2014solitary} to consider non-identical groups without
phase shift ${\a_a = \a_r = 0}$. The oscillators in each group are identical,
but there exists a difference in the natural frequencies $\w$, i.e.,  in
Eqs.~\eqref{eq:m_kuramoto_ode} ${\w_a = 0}$ and ${\w_r = \w}$. In this case the
solitary state changes: the cluster of the repulsive oscillator has a small
phase-shift in relation to the cluster of the attractive oscillators and the
phase-shift between the repulsive cluster (or the attractive cluster) and the
solitary oscillator is no longer $\pi$. The solitary state also has full
measure, it will always be reached, regardless of the initial condition.  The
region of existence of the solitary state, as well as the phase shifts between
the clusters and the solitary oscillator can be calculated and fits well to
numerical observations, see Fig.~\ref{fig:solitary_state}(b). The state is also
not stationary, but rotates with a constant frequency
\begin{equation}
  \n = \frac{1 + \e}{\e} \w \; .
  \label{eq:solitary_frequency}
\end{equation}

\begin{figure}
  \centering
  \includegraphics{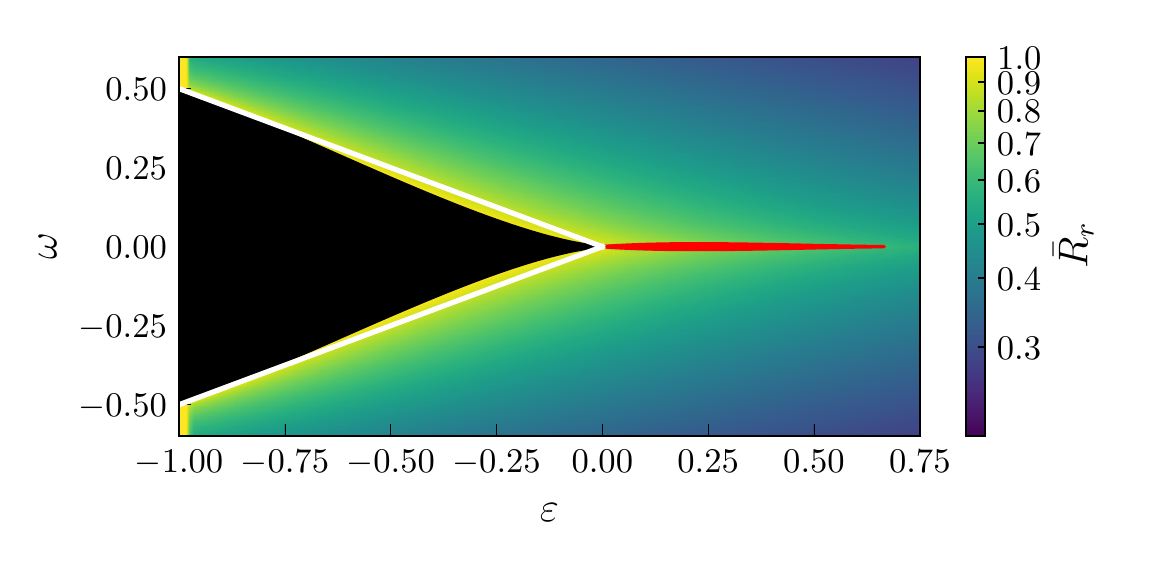}
  \caption{Parameter space of the M-Kuramoto model in
  Eqs.~\eqref{eq:m_kuramoto_ode} with non-identical groups. The color shows the
  average order parameter of the repulsive group $\bar{R}_r$. Full synchrony is
  denoted by the black region, the solitary state by the red region and the
  white lines show the condition of existence for the fully synchronous state.}
  \label{fig:m_kuramoto_overview}
\end{figure}

Aside from the solitary, there exist two other states in the system
(Fig.~\ref{fig:m_kuramoto_overview}), a fully synchronous state and a
self-consistent partial synchronous state. In the fully synchronous state both
clusters are fully synchronized, have a constant phase shift, and rotate with a
uniform frequency. The region of existence and the stability of the state can be
calculated directly from Eqs.~\eqref{eq:m_kuramoto_ode}. The results show that
the region of stability with
\begin{equation}
  \w = \pm \sqrt{- \frac{\e^3}{2} - \frac{\e^4}{4}}
\end{equation}
is slightly smaller than the region of existence, which fits the numerical
results in Fig.~\ref{fig:m_kuramoto_overview}. The frequency has the same
relation as the solitary state in Eq.~\eqref{eq:solitary_frequency}. We also
find numerically that the attractive group always fully synchronizes, even
outside the fully synchronous state, although there is no analytical
proof for this. A simple check of the stability for the attractive cluster
yields the inequality
\begin{equation}
  R_r \cos(\t_r - \t_a) < (1 + \e)^{-1} \; .
\end{equation}
The exact dynamics of the mean-field quantities are unknown, so this cannot be
solved analytically, but simulations show that ${\t_r - \t_a}$ is quite small
and $R_r$ falls off very quickly with $\e$ so that this equality is always
fulfilled, and the attractive group fully synchronizes.

\begin{figure}
  \centering
  \includegraphics{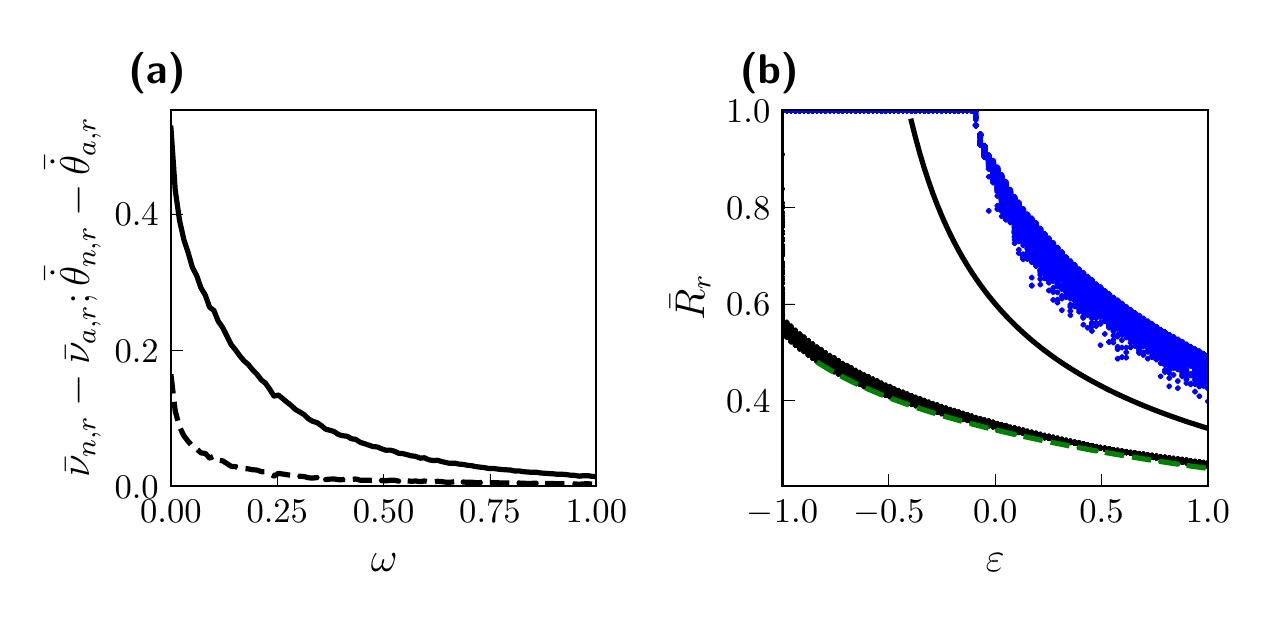}
  \caption{Goodness of fit for the WS theory in the M-Kuramoto model from
  Eqs.~\eqref{eq:m_kuramoto_ode}. In (a) the difference between the numerical
  ($\bar{\n}_{n, r}$) and analytical ($\bar{\n}_{a, r}$) average frequency for
  the single oscillators (dashed black line) and the mean-field (solid black
  line, $\bar{\dot{\t}}_{r}$) of the repulsive group is shown.  The frequencies
  were averaged over 100 different initial conditions for $N_a = N_r = 5$ and
  $\e = 0.5$. In (b) the average order parameter is plotted, where the blue
  crosses (black dots) show different initial conditions for $\w = 0.02$ ($\w =
  0.6$) and their analytical value as a solid black line (green dashed line).}
  \label{fig:partial_synchrony_ws}
\end{figure}

The final state, self-consistent partial synchrony, is defined by the difference
in average frequencies between the single oscillators and their
mean-field~\cite{clusella2016minimal}. The oscillators of the repulsive group
move at a frequency that is generally faster than their mean-field. The WS
theory cannot be applied to the whole system at once to explain this behavior,
but on each group separately~\cite{pikovsky2008partially}. Making the crude
approximation of the thermodynamic limit as well as stationarity (numerically we
find that the average frequencies of both mean-fields coincide), allows for the
calculation of the mean-field variables on the Ott-Antonsen manifold. As
expected, they do not fit well for small $\w$ or $\e$, but give a surprisingly
good approximation for big $\w$ and $\e$, even for a small system of ${N_a = N_r
= 5}$, see Fig.~\ref{fig:partial_synchrony_ws}. With the now  known average
values of $R_r$ and $\dot{\t}_r$ it is possible to calculate the average
frequency of the oscillators in the WS theory~\cite{baibolatov2009periodically}
as (bars denote the time-average)
\begin{equation}
  \bar{\n}_r = \frac{1 - \bar{R}_r^2}{1 + \bar{R}_r^2} \w
    + \frac{2 \bar{R}_r^2}{1 + \bar{R}_r^2} \bar{\dot{\t}}_r \; ,
\end{equation}
which shows clearly the expected difference between the average frequency of the
mean-field and the single oscillators.

Even in a simple model of phase coupled oscillators in the weak coupling limit,
it is possible to find interesting dynamical states. This can be extended
further by considering more complex coupling schemes than simple all-to-all
coupling. For even richer dynamics, higher-order coupling terms need to be
considered.

\subsection{Higher Order Phase Dynamics - Stuart-Landau Model}
\label{sec:sl}
Extending the first order phase approximation for coupled oscillators in
Eq.~\eqref{eq:phase_first_order} to higher orders needs knowledge of the phase
dynamics of the uncoupled units.  One of the oscillators with a known analytical
phase is the Stuart-Landau
oscillator~\cite{pikovsky2003synchronization,landau1944problem}.  Its
dimensionless dynamics in a coupled system are
\begin{equation}
  \frac{d A}{dt} = (1 + i \w) A - \abs{A}^2 A - i \g A(\abs{A}^2 - 1)
    + \e G(A_1, A_2, \ldots) \; ,
\end{equation}
where $A$ is a complex amplitude, $\w$ is the frequency and $\g$ is the
non-isochronicity parameter and determines $\Vp(A)$. Using ${A = R e^{i \t}}$
leads to
\begin{equation}
  \begin{aligned}
    \dot{R} & = R - R^3 + \e \Re{e^{-i \t} G} \\
    \dot{\t} & = \w - \g (R^2 - 1) + \e R^{-1} \Im{e^{-i \t} G} \; .
  \end{aligned}
  \label{eq:sl_polar}
\end{equation}
From there it follows for the phase of the uncoupled system  ${\vp = \t - \g
\ln(R)}$ and
\begin{equation}
  \begin{aligned}
    \dot{R} & = R - R^3 + \e \Re{e^{-i (\vp + \g \ln R)} G} \\
    \dot{\vp} & = \w + \e R^{-1}
      \left( \Im{e^{-i (\vp + \g \ln R)} G}
      - \g \Re{e^{-i (\vp + \g \ln R)} G} \right) \; .
  \end{aligned}
  \label{eq:sl_phase}
\end{equation}

In Ref.~\cite{matheny2019exotic}, eight nanomechanical systems (NEMS) with
dynamical equations resembling the Stuart-Landau oscillator were coupled in a
circle. The oscillators were connected with their two neighbors, and the
coupling term consisted of the average of them. The resulting phase reduction up
to the second order in the coupling strength contained terms not present as
physical links in the coupling scheme, such as ${\sin(\vp_{k+2} - \vp_k)}$,
${\sin(\vp_{k+2} - 2 \vp_{k+1} + \vp_{k})}$ and similar terms in the opposite
direction. These non-structural terms only appear in the second-order
approximation and would not be recovered in the typical first-order phase
reduction. The observed system exhibits many dynamical states, such as traveling
waves or weak chimeras, which would not necessarily be expected in the
first-order phase approximation. Numerical simulations of the phase model show
good agreement with the experimental results, pointing to the importance of the
non-structural coupling terms in complex synchronization behavior.

Similarly, in Ref.~\cite{leon2019phase} a system of coupled Stuart-Landau
oscillators was investigated numerically. The oscillators were coupled
all-to-all to their mean-field. In the phase reduction up to the second-order in
$\e$ appeared terms of the form ${\sin(\vp_m + \vp_n - 2\vp_k)}$, where $m$, $n$
and $k$ are all possible combinations, coupling three oscillators. They
determined that these terms are necessary to explain the system's full dynamics
but are again not present in the first-order phase dynamics, which only
contains the existing pairwise connections.

In both Refs~\cite{leon2019phase,matheny2019exotic} with coupled
Stuart-Landau-like oscillators, the first-order approximation of the phase
dynamics yields a Kuramoto-like model. This allows the novel terms to be seen as
an extension of the Kuramoto model in Eq.~\eqref{eq:kuramoto} to higher coupling
strength.

\begin{figure}
  \centering
  \includegraphics{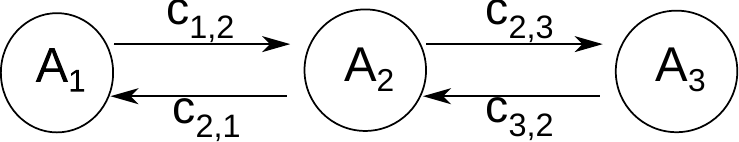}
  \caption{A system of three Stuart-Landau oscillators coupled in a line. The
  $A_k$ denote the complex amplitudes and the $c_{j,k}$ the coupling terms
  between the oscillators.}
  \label{fig:sl_coupling_scheme}
\end{figure}

Based on the observations for the NEMS we investigated the higher order phase
reduction for a system of three coupled nonidentical Stuart-Landau oscillators
in Ref.~\cite{gengel2020high}. In difference to Ref.~\cite{leon2019phase} one
pairwise connection is missing. The coupling scheme is a line, as can be seen in
Fig.~\ref{fig:sl_coupling_scheme} and is given by
\begin{equation}
  G_k(A_k, A_{k-1}, A_{k+1}) = c_{k-1, k} e^{i\b_{k-1, k}} A_{k-1}
    + c_{k+1, k} e^{i\b_{k+1, k}} A_{k+1} \; ,
  \label{eq:sl_coupling_function}
\end{equation}
where the $c_{j, k}$ are of ${\mathcal{O}(1)}$ and the $\b_{j, k}$ are phase
shifts between the two oscillators.  To extend the phase reduction beyond the
first order in $\e$ we use a perturbation method, where the $R$ and $\dot{\vp}$
are functions of the phases and expand them as a power series
\begin{equation}
  \begin{aligned}
    R & = 1 + \e r^{(1)}(\vp, \vp_1, \vp_2, \ldots)
      + \e^2 r^{(2)}(\vp, \vp_1, \vp_2, \ldots) + \ldots \\
    \dot{\vp} & = \w + \e \p^{(1)}(\vp, \vp_1, \vp_2, \ldots)
      + \e^2 \p^{(2)}(\vp, \vp_1, \vp_2, \ldots) + \ldots \; .
  \end{aligned}
  \label{eq:phase_expansion}
\end{equation}
In the dimensionless form the limit cycle of the uncoupled oscillator has the
amplitude ${R_0 = 1}$, as can be easily checked in Eq.~\eqref{eq:sl_polar}.
Inserting these assumptions in Eqs.~\eqref{eq:sl_phase}, the dynamics of
$r^{(1)}$, $r^{(2)}$, $\p^{(1)}$, $\p^{(2)}$ and so on can be found by gathering
the powers of $\e$. The full calculation is rather long, so please see
Ref.~\cite{gengel2020high} for a detailed explanation. Here it will suffice to
say that to find the phase dynamics in the  second order in $\e$, a partial
differential equation has to be solved using a Fourier series to represent the
coupling function $G_k$. This representation is motivated by the fact that $G_k$
has to be $2\pi$-periodic for all the phases. The resulting reduction for the
first oscillator up to the second order in $\e$ is
\begin{equation}
  \begin{aligned}
    \dot{\vp}_1 & = \w_1 + \e c_{2,1} [\sin(\vp_2 - \vp_1 + \b_{2,1})
        - \g \cos(\vp_2 - \vp_1 + \b_{2,1})] \\
      & + \e^2 \left[ a_{1; 0,0,0}^{(2)}
        + a_{1; -2,2,0}^{(2)}\cos(2\vp_2 - 2\vp_1)
        + b_{1; -2,2,0}^{(2)}\sin(2\vp_2 - 2\vp_1) \right. \\
      &  \left. + a_{1; -1,2,-1}^{(2)}\cos(2\vp_2 - \vp_1 -\vp_3)
        + b_{1; -1,2,-1}^{(2)}\sin(2\vp_2 - \vp_1 - \vp_3) \right. \\
      & \left. + a_{1; -1,0,1}^{(2)}\cos(\vp_3 - \vp_1)
        + b_{1; -1,0,1}^{(2)}\sin(\vp_3 - \vp_1)\right] \; ,
   \end{aligned}
   \label{eq:sl_first_phase_dynamics}
\end{equation}
with $a_{k; \vec{l}}^{(j)}$ and $b_{k; \vec{l}}^{(j)}$ denoting the coefficients
for cosine and sine terms respectively and $k$ being the index of the
oscillator. A vector of integers $\vec{l}$ contains the coefficients before the
phases and $j$ the power in $\e$.

Using the perturbation method up to the second order in $\e$ yields again
non-structural terms in Eq.~\eqref{eq:sl_first_phase_dynamics}, e.g. terms of
the form ${\sin(\vp_3 - \vp_1)}$. Aside from these, there exist additional terms
coupling all three oscillators ${\sin(2\vp_2 - \vp_1 -\vp_3)}$ and second
harmonics ${\sin(2\vp_2 - 2\vp_1)}$. The coefficients of the second order also
contain the frequency difference between the oscillators, whereas the first
order terms do not.

A Fourier Ansatz is used to measure the coupling terms numerically and verify
the analytical results. The phase dynamics are ${2\pi}$-periodic, so they can be
written as a multidimensional Fourier series
\begin{equation}
  \dot{\vp}_k = a_{k; \vec{0}}
    + \sum_{\vec{l} \neq 0} \left[ a_{k; \vec{l}} \cos(\vec{\vp} \cdot \vec{l})
    + b_{k; \vec{l}} \sin(\vec{\vp} \cdot \vec{l}) \right] \; ,
  \label{eq:phase_fourier}
\end{equation}
where ${\vec{\vp} = (\vp_1, \vp_2, \ldots, \vp_N)}$ is a vector of all the
phases, $\vec{l}$ an N-dimensional vector of integers and ${\vec{\vp} \cdot
\vec{l} = \sum_j \vp_j l_j}$ the scalar product. The $a_{k,
\vec{l}}$ and $b_{k, \vec{l}}$ are Fourier coefficients. This form resembles
the analytical results in Eq.~\eqref{eq:sl_first_phase_dynamics} and finding the
relevant coupling terms is reduced to fitting the Fourier coefficients
$a_{k; \vec{l}}$ and $b_{k; \vec{l}}$ up to some maximum value with ${\abs{l_j}
\leq m}$. Because the coupling function is real, ${a_{k;-\vec{l}} = - a_{k;
\vec{l}}}$ and the number of coefficients to consider halves. Still, the number
of terms that need to be fitted scales like $m^3$, which increases the necessary
number of data points for a good fit very rapidly with $m$ and leads to the
curse of dimensionality.

In the case of partial synchrony, the Fourier coefficients can be fitted
onto one long time series, although some initial transient has to be integrated
over before the dynamics reach the torus spanned by the phases. For a good fit,
the trajectory should be long enough to cover the whole torus. However, this is
not the case in the synchronous regime; instead, the dynamics will settle on a
single synchronized trajectory. The integration has to be stopped before
reaching this synchronous trajectory and restarted with different initial
conditions until the torus is sufficiently filled.

\begin{figure}
  \centering
  \includegraphics{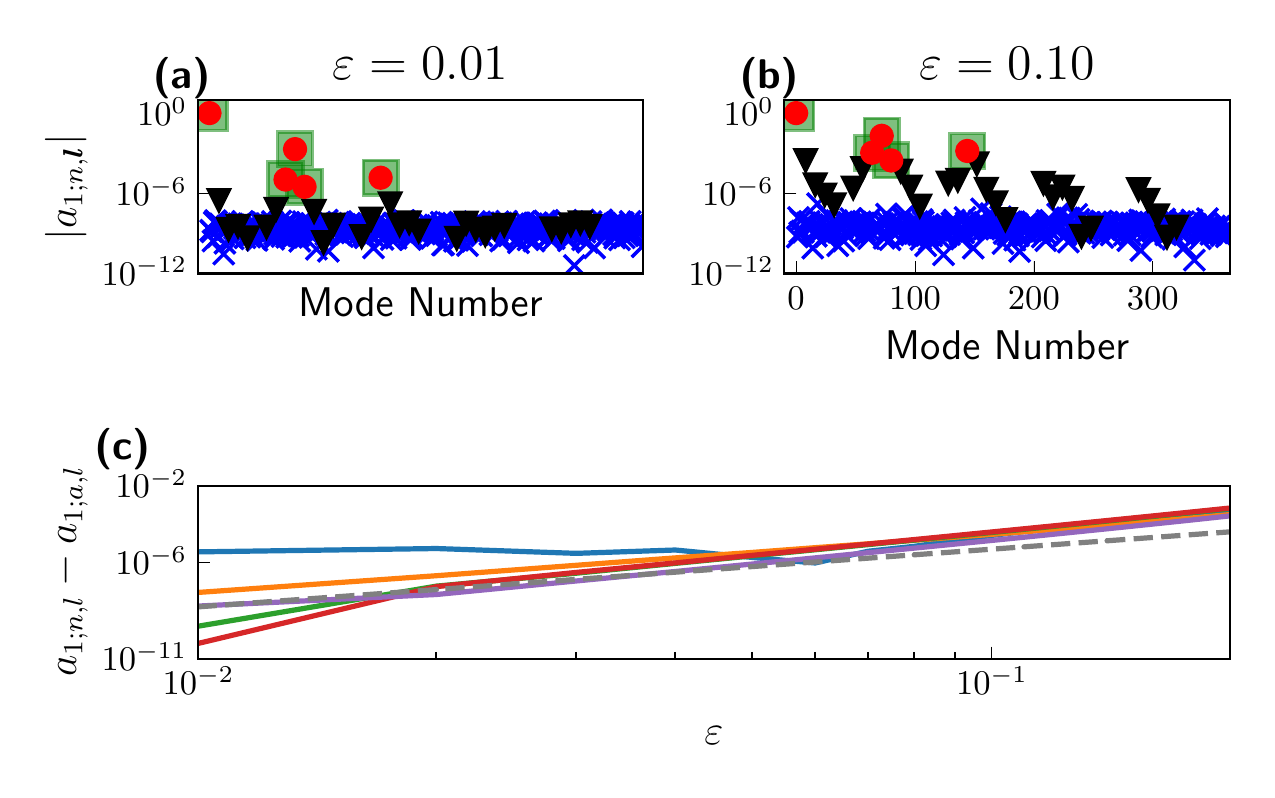}
  \caption{Comparison of analytical and numerically fitted modes for the
  Stuart-Landau oscillators with coupling function
  Eq.~\eqref{eq:sl_coupling_function} in the partial synchronous regime. The
  modes shown are the cosine terms for the first oscillator in
  Eq.~\eqref{eq:sl_first_phase_dynamics}. In (a) and (b) the red circles denote
  numerically fitted values for terms appearing in
  Eq.~\eqref{eq:sl_first_phase_dynamics}, while their analytical values are
  shown as green boxes. Black triangles denote terms ${\sum_j l_j = 0}$, which
  are the only allowed terms, because of rotational symmetry. The blue crosses
  are all other modes up to ${\vec{l} = (4, 4, 4)}$.  In (c) the difference
  between the analytical and numerically fitted modes are shown. The gray dashed
  line is a polynomial $\e^3$, so the error scales stronger than $\e^3$.}
  \label{fig:sl_phase_fit}
\end{figure}

Fitting the coefficients for different coupling strengths $\e$ yields the power
series
\begin{equation}
  a_{k; \vec{l}} = a_{k; \vec{l}}^{(0)} + \e a_{k; \vec{l}}^{(1)}
    + \e^2 a_{k; \vec{l}}^{(2)} + \ldots
\end{equation}
and a similar series for $b_{k; \vec{l}}$. The coupling function is then
reconstructed by comparing these series and Eq.~\eqref{eq:phase_fourier}. The
results of this method fit very well to the analytical prediction, as can be
seen in Fig.~\ref{fig:sl_phase_fit}.

The numerical method to reconstruct the phase can also be used in cases where
the phase is unknown, but then the phases and their derivatives have to be
calculated numerically. For finding the phase, an autonomous copy of each
oscillator is integrated. It evolves for a number of its autonomous periods $T$
until it reaches its limit cycle. The phase after the relaxation is then also
the phase of the point before the relaxation. To find the derivative, one
observes the infinitesimal time step $dt$ of the perturbed system and sets it in
relation to a different time step $\overline{dt}$ on the limit cycle. This time
difference $\overline{dt}$ is determined by the motion on the limit cycle and
the derivative $\vec{f}$ in Eq.~\eqref{eq:dynamical_system}.  Using the
relation of $dt$ and $\overline{dt}$ allows the calculation of the phase
derivative, even if the oscillator is perturbed far from its limit cycle.  For a
full explanation and the resulting equation, see Ref.~\cite{gengel2020high}.

\section{Summary}
Phase dynamics are an important tool to analyze dynamical systems. In the case
of a simple perturbation of an ensemble of oscillators, this reduces, in the
first order, to the collective PRC. The PRC has been investigated for a system
of coupled Rayleigh oscillators, where it took a long time to fully relax back
onto the limit cycle after the perturbation, even in comparison to the time
needed for synchronization. This makes the collective PRC only usable as an
approximate description in noisy environments, where it is perturbed again
before it can fully settle.

In the paradigmatical Kuramoto model with two groups, one attractive and one
repulsive, an interesting solitary state is observed that does not appear in a
model with just one group.  A single solitary oscillator leaves its otherwise
fully synchronized group in this state and gets phase-shifted by $\pi$. The use
of groups with different natural frequencies leads to a stabilization of this
state.

In the case of phase dynamics for the description of a coupled system, most of
the current works use the first-order phase approximation.  The first-order
reduction for pairwise coupled units yields only pairwise terms in the phase
model. When extending the phase reduction beyond the first-order, new
connections arise that are necessary to describe more complicated dynamics. Even
in a simple model of three Stuart-Landau oscillators, coupled in a line, the
second-order approximation consists only of higher-order or non-structural
terms. The analytical derivation of the additional terms follows from a simple
perturbation Ansatz. A numerical verification shows good agreement and supports
the analytical findings.

\begin{acknowledgement}
The author thanks Michael Rosenblum for his advice.  This paper was developed
within the scope of the IRTG 1740 / TRP 2015/50122-0, funded by the DFG /
FAPESP.

The author confirms the sole responsibility for the following: study conception
and design, data collection, analysis and interpretation of the results,
literature selection and manuscript preparation.
\end{acknowledgement}

\end{document}